\documentclass[useAMS,usenatbib,onecolumn]{mn2e}
\usepackage{graphics,epsfig}

\title{On the theory of astronomical maser. II. Polarization of maser radiation}

\author[Dinh-V-Trung]{Dinh-V-Trung\thanks{on leave from Center for Quantum Electronics, Institute of Physics,
10 DaoTan Street, BaDinh, Hanoi, Vietnam}\\
Institute of Astronomy and Astrophysics, Academia Sinica\\ 
P.O Box 23-141, Taipei 10617, Taiwan\\
email:trung@asiaa.sinica.edu.tw}
\begin{document}
\date{}
\maketitle
\begin{abstract}
In this paper we investigate the polarization property of the radiation amplified by 
astronomical masers in the presence of a strong magnetic field. 
Our model explicitly takes into account the broadband nature of the radiation field
and the interaction of the radiation with the maser transition $J=1-$0.  
The amplification of different realisations of the background continuum radition 
by the maser is directly simulated and the Stokes parameters of the radiation field 
are then obtained by averaging over the ensemble of emerging maser radiation. For isotropic
pumping and partially saturated masers we find that the maser radiation is linearly polarized
in two representative cases where the magnetic field {\bf B} makes an angle $\theta$=30$^0$ and $\theta$=90$^0$
to the maser axis. The linear polarization for maser radiation obtained in our simulations for both cases 
are in agreement with the
results of the standard model. Furthermore, no instability during amplification is seen in
our simulations. Therefore, we conclude that there is no problem with the previous 
numerical investigations of maser polarization in the unsaturated and partially saturated regime. 
\end{abstract}

\section{Introduction}
Strong maser radiation has been detected in various astronomical environments, from 
nearby starforming regions to distant galactic nuclei. The extremely high brightness temperature 
of astronomical masers allows very high
angular resolution observations to study the kinematics and physical properties of the
masing environments (Reid \& Moran 1981). It has been recognized that maser radiation is
generally polarized, either linearly or circularly. In some cases the polarization
degrees up to 100\% have been observed. Because polarization of the radiation field is related to
spatial anisotropy, i.e the presence of a strong and ordered magnetic field, maser radiation
holds the promise of providing valuable information on the elusive magnetic field.\\
A complete understanding of the mechanism to generate polarization of maser emission is essential
to interpret observational results and to infer the physical properties, such as magnetic field 
direction and strength, in the maser environment. Theoretical study of astronomical masers started
with the work of Litvak (1970). Subsequently Goldreich et al. (1973) worked out the basic model
of maser polarization for the J=1--0 transition. Their work predicts that, in the presence of a strong
magnetic field, radiation from saturated masers is linearly polarized, reaching a limit of 
Q/I=$-$1 for $sin^2\,\theta\,\le\, \frac{1}{3}$ and (3$sin^2\,\theta - 2$)/3$sin^2\,\theta$ for 
$sin^2\,\theta\,\ge \, \frac{1}{3}$. However,
the result is derived only for the case of a fully saturated maser and the question of how that
limiting polarization is reached for any real astronomical maser remains open.
Numerical studies by Western \& Watson (1984), Deguchi \& Watson (1990) and Nedoluha \& Watson (1990)
provide the dependence of fractional linear polarization on the saturation parameter R/$\Gamma$, the ratio between
the stimulated emission rate R and the loss rate $\Gamma$. Numerical results show that for the
J=1--0 transition, the linear polarization increases slowly toward the limiting solution, and for higher
lying rotational transitions such as J=2--1, the limiting polarization predicted by Goldreich et al. (1973) 
can only be attained in the fully saturated regime, namely R/$\Gamma$ $\sim$ 100 or even higher. 
Observationally, such high intensities are almost impossible to be
realised in astronomical masers. More detailed calculations by Deguchi \& Watson (1990), taking
into account population mixing between magnetic sub-levels in the regime of large stimulated emission rate,
actually predict that the polarization fraction decreases and disappears at high enough intensity. Thus, the
limiting polarization can never be reached in high lying transitions such as J=2--1.\\
Observationally, SiO maser lines in vibrationally excited state $\upsilon=1$, such as J=3--2 and even J=5--4, 
from evolved stars have been known to possess large linear polarization (McIntosh \& Predmore 1993).
This discrepancy is often cited (Elitzur 1993, 1995) as deficiency of the framework under 
which previous numerical studies are carried out. Recently, in a series of papers (Elitzur 1991, 1993, 1996) a 
different model of the maser polarization was proposed. 
The linear polarization fraction is determined through finding the eigenvalues of the radiation transfer equations. 
In an important departure from previous studies, an ensemble average of the different modes of the radiation field, which
is inherently broadband and random, is considered. Some novel solutions are found, namely that
unsaturated maser emission can have the
same polarization fraction as saturated masers and when the magnetic field is close to the maser axis
($\sin^2\theta\,\le\,\frac{1}{3}$), propagation of polarized emission is forbidden. In addition,
the same analysis predicts that polarization properties of astronomical masers are spin independent, i.e
high lying transitions should behave in the same way as the often studied J=1--0 transition.
Instability during 
the amplification of maser radiation is cited as the main reason for the new solutions, although in 
subsequent publications (Elitzur 1995, 1996), 
more emphasis is put on the ensemble average over modes or configurations of the radiation field. 
Elitzur (1995) points out that a full understanding of the creation and evolution of polarization in masers
requires simulations involving a statistical ensemble of waves. These new results are 
vehemently criticized by Watson (1994), who performs again the stability analysis on 
the radiation transfer equations and finds no instability. More recently, Gray (2003) points that
the multi-level model (Field \& Gray 1988, Gray \& Field 1995) can be formally reduced to the
idealized two-level case with a similar set of equations as in Watson (1994), 
leading to the same predictions for the linear polarization properties
of astronomical masers as obtained earlier by Watson and co-workers.\\ 
Obviously, the explicit incorporation of the broadband random radiation field into the study of astronomical 
masers is of geat importance. That kind of simulation might help to elucidate the current debate regarding 
the polarization theory of astronomical masers.
To simulate from first principles the amplification of radiation by a masing medium, as suggested by 
Elitzur (1995), is a complicated task. 
In addition to a new formulation of the interaction process between
radiation field and the maser medium, a dramatic increase in computing power to follow the evolution of a large number
of realisations of the incident radiation field is also required.\\ 
So far the only work aimed to treat the broadband maser radiation from first principles, and in a 
transparent and self-consistent manner, is published in Menegozzi \& Lamb (1978). However, due
to limited computing power, they could not perform a large enough number of simulations to effectively draw any
firm conclusion on the statistics of the maser radiation field.
In a previous paper (Dinh-V-Trung 2009) we have followed the formulation of Menegozzi \& Lamb (1978) to investigate
the standard theory of a scalar maser and the statistics of the maser radiation field.  
In this paper we generalize our model of the astronomical maser to include the vector 
nature of the radiation field. That will allow us perform simulations of polarized maser 
radiation, which is the main focus of this paper.
We hope that our work will provide a small step toward a better understanding of the 
properties of astronomical masers.
\section{Basic theory}
\subsection{Radiation field}
In this paper we consider a one-dimensional maser and assume that a strong magnetic field exists inside 
the maser medium and makes an angle $\theta$ to the
propagation direction of the radiation. That means the Zeeman splitting 
$\mathit{g}\omega_{\rm B}$ between magnetic 
sub-levels of a particular rotational level J is much greater than 
the decay rates due to collisions and/or radiative transitions. As a result, 
the magnetic field defines a preferred direction in space and a good quantization axis.
We adopt here the geometry used by Deguchi \& Watson (1990) in which the magnetic field 
$\mathbf{B}$ is aligned with the Oz$^\prime$
axis of the {\bf{B}}-frame. The {\bf{k}}-frame which has the Oz axis directed 
along the propagation direction is obtained from the {\bf{B}}-frame system by 
rotating through an angle $\theta$ about the Ox$^\prime$ axis.\\
The spherical basis is defined in the conventional way (Brink \& Satchler 1994, Zare 1988):
\begin{eqnarray}
\mathbf{e}_{\rm p=\pm 1} & = & \mp\left(\mathbf{e}_{\rm x} + 
\mathbf{e}_{\rm y}\right)/\sqrt{2}; \nonumber \\
\mathbf{e}_{\rm p=0} & = & \mathbf{e}_{\rm z} 
\end{eqnarray}
where ($\mathbf{e_{\rm x}},\mathbf{e_{\rm y}},\mathbf{e_{\rm z}}$) are the unit vectors along
the frame axis.
Similarly the spherical bases in the {\bf{B}}-frame is defined as:
\begin{eqnarray}
\mathbf{e}_{\rm M=\pm 1} & = & \mp\left(\mathbf{e}_{\rm x'} + 
\mathbf{e}_{\rm y'}\right)/\sqrt{2} \nonumber \\
\mathbf{e}_{\rm M=0} & = & \mathbf{e}_{\rm z'} 
\end{eqnarray} 
The relation between the two bases associated with the {\bf{k}}-frame and the {\bf{B}}-frame
follows easily:
\begin{eqnarray}
\mathbf{e}_{\rm p=\pm 1} & = & \mp(\mathbf{e}_{\rm x'}\, 
\pm \, i\, {\rm cos}\theta\,
\mathbf{e}_{\rm y'}\, \pm\,i\,{\rm sin}\theta\,\mathbf{e}_{\rm z'})/\sqrt{2} \nonumber \\
\mathbf{e}_{\rm p=0} & = & {\rm sin}\theta\,\mathbf{e}_{\rm y'}\, 
+ \, {\rm cos}\theta\,\mathbf{e}_{\rm z'}
\end{eqnarray}
We represent the electric field of the maser radiation and the induced macroscopic 
polarization vector in the spherical bases:
\begin{eqnarray}
\mathbf{E}{\rm (z,t)} & = & Re\left[\sum\limits_{p=\pm}^{}\, (-1)^{\rm p} {\rm E}_{\rm p}
\mathbf{e}_{\rm -p} \right] \nonumber \\
\mathbf{P}{\rm (z,t)} & = & Re\left[\sum\limits_{p=\pm}^{}\, (-1)^{\rm p} {\rm P}_{\rm p}
\mathbf{e}_{\rm -p} \right]
\end{eqnarray}
Where ${\rm E}_{\rm p}$ and ${\rm P}_{\rm p}$ are spherical components of the 
electric field and polarization vector, respectively.\\
Because our aim is to simulate the spectral properties of the maser radiation, it is preferable to work from the
beginning in the frequency domain.
In expressing the frequency dependence of the electric field and polarization
vector, we retain only the positive frequencies. The anti-resonant (negative) frequencies are ignored.
This approximation is usually referred to as the rotating wave approximation. The spherical
components of the electric field and polarization vector can be written as: 
\begin{eqnarray}
{\rm E}_{\rm p}(z,t) & = & E_{\rm p}(z,t)\, e^{-i\omega_{0} (t\, - \, z/c)} \nonumber \\
{\rm P}_{\rm p}(z,t) & = & P_{\rm p}(z,t)\, e^{-i\omega_{0} (t\, - \, z/c)}
\end{eqnarray}
Where $\omega_{0}\,=\,2\pi\nu_0$ is the angular frequency at the maser line center. 
The amplitudes $E_{\rm p}(z,t)$ and $P_{\rm p}(z,t)$ are 
assumed to vary slowly with time in comparison
to the term $e^{-i\omega_{0}\,t}$. Using the spectral representation theorem (Priestley 1981), for any realisation of the radiation
field during the time 
interval $T$ the amplitude of electric field and polarization vector can be expressed in terms of Fourier series:
\begin{eqnarray}
E_{\rm p}(z,t) & = & \sum_{\rm n=-\infty}^{\rm n=+\infty}E_{\rm p}(z,\omega_n)\, 
e^{-i\omega_n(t\, - \,z/c)} \nonumber \\
P_{\rm p}(z,t) & = & \sum_{\rm n=-\infty}^{\rm n=+\infty}P_{\rm p}(z,\omega_n)\, 
e^{-i\omega_n(t\, - \,z/c)}
\end{eqnarray}
where $\omega_n$ = $2\pi n/T$.
The Stokes parameters of the radiation field are the quantities measured directly in observations of astronomical masers. These
parameters are commonly used to characterize the polarization properties of the maser radiation. Because the
radiation field is stationary and ergodic, these parameters can be defined as either an ensemble average
or time average of different realisations of the radiation field. For the convenience of presenting the 
simulation results, we will follow the same convention as in Deguchi \& Watson (1990) in defining the
four parameters $({\cal I,Q,U,V})$ for each frequency $\omega_n$ during an interval $T$ of each realisation:
\begin{eqnarray}
{\cal I}(\omega_n)\Delta\omega & = & \frac{c}{8\pi}[E_{-}(\omega_n)E_{-}^*(\omega_n) + E_{+}(\omega_n)E_{+}^*(\omega_n)] 
\nonumber \\
{\cal V}(\omega_n)\Delta\omega & = & \frac{c}{8\pi}[E_{-}(\omega_n)E_{-}^*(\omega_n) - E_{+}(\omega_n)E_{+}^*(\omega_n)] \\
\left[{\cal Q}(\omega_n)\,-\,i{\cal U}(\omega_n)\right] \Delta\omega & = & \frac{c}{4\pi}\left[E_{-}(\omega_n)E_{+}^*(\omega_n)
\right] \nonumber
\end{eqnarray}
In this definition, all the parameters $({\cal I,Q,U,V})$ are real, for example ${\cal Q}$ and ${\cal U}$ can be obtained directly from 
the real and imaginary part of the last equation.
The inclusion of $\Delta\omega$=$2\pi/T$ is due to the fact that we use Fourier series to decompose the radiation
field into discrete harmonic components. Each Fourier component represents the radiation field in the
frequency band of $\Delta\omega$. The coherence time of the electric field (Mandel \& Wolf 1965) is
indeed the interval $T$. Thus, during this interval of time, the radiation field within the frequency band $\Delta\omega$ 
can be considered as 
quasi-monochromatic. The four parameters $({\cal I,Q,U,V})$, which are real and defined as above, are similar 
to Stokes parameters of a monochromatic wave and statisfy the usual relation:
\begin{equation}
{\cal I}^2(\omega_n) = {\cal Q}^2(\omega_n) + {\cal U}^2(\omega_n) + {\cal V}^2(\omega_n)
\end{equation}
We emphasize here that these parameters contain all the information on the amplitudes and the
relative phase between different spherical components ($E_{+}$ and $E_{-}$) of the radiation field in each
frequency band. 
The usual Stokes parameters of the random radiation field are then the ensemble averages, denoted as $\left<...\right>$, 
of the above $({\cal I,Q,U,V})$ parameters:
\begin{eqnarray}
I(\omega_n) & = &  \left<\, {\cal I}(\omega_n)\,\right> \nonumber \\
Q(\omega_n) & = &  \left<\, {\cal Q}(\omega_n)\,\right> \nonumber \\
U(\omega_n) & = &  \left<\, {\cal U}(\omega_n)\,\right>  \\
V(\omega_n) & = &  \left<\, {\cal V}(\omega_n)\,\right> \nonumber
\end{eqnarray}
For unpolarized continuum radiation, the harmonic components of different polarization and 
frequency are independent random variables having zero mean and the same variance. 
Therefore, after taking the ensemble average, the ensemble-averaged Stokes parameters $Q(\omega_n)$, 
$U(\omega_n)$ and $V(\omega_n)$ vanish. We note that the information on the
amplitude and relative phase of different spherical components of the radiation field at each frequency band is
lost after the step of taking the ensemble average. That is the fundamental difference between the four parameters
$({\cal I,Q,U,V})$ defined above for each realisation and the Stokes parameters.
\subsection{Radiation-matter interaction}
The basic transfer equations of maser radiation are derived here for the case involving the transition J=1--0. 
For higher transitions the procedure is similar but algebraic manipulation is considerably more involved. 
Because we consider a one-dimensional maser amplifying the background continuum radiation, the spontaneous
emission of masing molecules is ignored. This assumption is often used in theoretical studies of astronomical
masers. To study self-consistently the effect of spontaneous emission, full quantum treatment of the radiation
field is necessary and will be the subject of future publications. Goldreich et al. (1973),
Deguchi \& Watson (1990) derived the transfer equation for the radiation field within the framework of
the rotating-wave approximation.
The transfer equation (Goldreich et al. 1973, Deguchi \& Watson 1990) can be written as follows:
\begin{equation}
\left(\frac{1}{c}\frac{\partial}{\partial\, t} \, +\, \frac{\partial}{\partial\, z}\right)\,E_{\rm p}(z,t) =
\frac{2\pi i \omega_{0}}{c}P_{\rm p}(z,t)
\end{equation}
or in the frequency domain:
\begin{equation}
\frac{d}{dz}E_{\rm p}(z,\omega) = \frac{2\pi i \omega_{0}}{c}P_{\rm p}(z,\omega)
\end{equation}
The polarization vector $\mathbf{P}$ describes how the masing medium interacts with the radiation field at the frequency of the
maser line. In this paper we will consider the case where the interaction between the radiation field and the molecule 
can be described within an electric dipole approximation. 
To calculate the value of the polarization vector $\mathbf{P}$ we need to use the density matrix $\rho(z, \upsilon, t)$ to describe the medium, 
which is a collection of masing molecules at a given position $z$ and moving at velocity $\upsilon$. 
The density matrix can be written as:
\begin{equation}
\rho(z,\upsilon,t)  =  \left( \begin{array}{llll}
\rho_{\rm ++} & 0 & 0 & \rho_{\rm +b} \\
0 & \rho_{\rm 00} & 0 & \rho_{\rm 0b} \\
0 & 0 & \rho_{\rm --} & \rho_{\rm -b} \\
\rho_{\rm b+} & \rho_{\rm b0} & \rho_{\rm b-} & \rho_{\rm bb} \\
\end{array}
\right)
\end{equation} 
The Hermitian property of the density matrix implies that $\rho_{\rm ba} = \rho_{\rm ab}^{\rm *}$
where ${\rm a} = \pm, 0$. By writing down the above form of the density matrix we have made the
assumption that off-diagonal elements $\rho_{\rm aa'}$ (a$\neq$a') are very small, which implies that
Zeeman splitting is much larger than the stimulated emission rates and thus the mixing of population of different
magnetic sub-levels does not occur (Goldreich et al. 1973).\\
To work in the frequency domain, we first expand the density matrix elements into Fourier series
for a given time interval $T$:
\begin{eqnarray}
\rho_{\rm ab}(z,\upsilon,t) & = & e^{-i\omega_0 (t\,-\, z/c)} \sum_{\rm n=-\infty}^{n=+\infty} 
[\rho_{\rm ab}(z,\upsilon,\omega_n) e^{-i\omega_n(t\,-\,z/c)}] \nonumber \\
\rho_{\rm aa}(z,\upsilon,t) & = & \sum_{\rm i=-\infty}^{i=+\infty} \rho_{\rm aa}(z,\upsilon,\omega_i) e^{-i\omega_i (t\,-\, z/c)} \\
\rho_{\rm bb}(z,\upsilon,t) & = & \sum_{\rm i=-\infty}^{i=+\infty} \rho_{\rm bb}(z,\upsilon,\omega_i) e^{-i\omega_i (t\,-\, z/c)} \nonumber
\end{eqnarray}
Because the diagonal elements of the density matrix $\rho_{\rm aa}$ and $\rho_{\rm bb}$ are real, we have the following
relations $\rho_{\rm aa}(z,\upsilon,\omega)$=$\rho_{\rm aa}(z,\upsilon,-\omega)^{\rm *}$ and similarly 
for $\rho_{\rm bb}(z,\upsilon,\omega)$.
The evolution equation of the density matrix $\rho(z,\upsilon,t)$ can be written in the 
following compact form (Icsevgi \& Lamb 1969, Sargent et al. 1974):
\begin{eqnarray}
\left( \frac{\partial}{\partial t} + \upsilon \frac{\partial}{\partial z} \right) \rho_{\rm ab}& = &
-(i\omega_{ab} + \Gamma)\rho_{\rm ab} - \frac{i}{\hbar}\mathsf{V}_{\rm ab}(\rho_{\rm bb} - \rho_{\rm
aa}) \nonumber \\
\left( \frac{\partial}{\partial t} + \upsilon \frac{\partial}{\partial z} \right) \rho_{\rm aa}& = &
\lambda_{\rm a} - \Gamma_{\rm a} \, \rho_{\rm aa} - \frac{i}{\hbar}(\mathsf{V}_{\rm ab}\rho_{\rm ba} - 
\mathsf{V}_{\rm ba}\rho_{\rm ab}) \\
 \left( \frac{\partial}{\partial t} + \upsilon \frac{\partial}{\partial z} \right) \rho_{\rm bb}& = &
\lambda_{\rm b} - \Gamma_{\rm b} \, \rho_{\rm bb} - \frac{i}{\hbar}(\mathsf{V}_{\rm ba}\rho_{\rm ab} - 
\mathsf{V}_{\rm ab}\rho_{\rm ba}) \nonumber
\end{eqnarray}
Where $\omega_{\rm ab}\,=\,\omega_0\,+\,{\rm a}\cdot g\omega_B$ are the frequencies of Zeeman components of the
$J=1-0$ transition, 
$\lambda_{\rm a}(\upsilon)$ and $\lambda_{\rm b}(\upsilon)$ are 
the pumping rates into the upper and lower maser levels. $\Gamma_a$ and $\Gamma_b$ are the loss rate due to 
pumping and collisional decoherence (Sargent et al. 1974). 
For the sake of simplicity, we assume here that the loss rates are the same for lower and upper levels of the maser transition. 
$\upsilon$ is the velocity of masing molecules. Substituting the Fourier expansion of density matrix elements into
the above equations (except for the terms involving the interaction matrix $\mathsf{V}$ to be written out explicitly later) and 
collecting term by term, we obtain:
\begin{eqnarray}
\rho_{\rm ab}(\omega_n, \upsilon) & = & -\frac{i}{\hbar}[{\rm V}_{\rm ab}(\rho_{\rm bb}\,-\,\rho_{\rm aa})(
\omega_n)]\cdot\gamma_{+}^{\rm ab}(\omega_n,\upsilon) \nonumber \\
\rho_{\rm aa}(\omega_n, \upsilon) & = & \left\{ \lambda_{\rm a}(\upsilon)\delta_{\rm n,0} - 
\frac{i}{\hbar}[\mathsf{V}_{\rm ab}\rho_{\rm ba} 
- \mathsf{V}_{\rm ba}\rho_{\rm ab})(\omega_n)]\right\}\cdot  \gamma_{+}^{\rm aa}(\omega_n,\upsilon) \\
\rho_{\rm bb}(\omega_n, \upsilon) & = & \left\{ \lambda_{\rm b}(\upsilon)\delta_{\rm n,0} - 
\frac{i}{\hbar}[\mathsf{V}_{\rm ba}\rho_{\rm ab} 
- \mathsf{V}_{\rm ab}\rho_{\rm ba})(\omega_n)]\right\}\cdot  \gamma_{+}^{\rm bb}(\omega_n,\upsilon) \nonumber
\end{eqnarray}
The $\gamma$ functions are the Lorentzian response of the masing molecules to the radiation
field and given as follows:
\begin{eqnarray}
\gamma_{\pm}^{ab}(\omega_n,\upsilon) & = & 1/\left\{\Gamma\: \pm \: i\,[\omega_{\rm ab} - (\omega_0\,+\,\omega_n)
\cdot(1\,-\,\frac{\upsilon}{c})] \right\} \,
\simeq \, 1/\left\{\Gamma\: \pm \: i\,[\omega_0\:\frac{\upsilon}{c}\,-\,(\omega_0\,+\,\omega_n\,-\,\omega_{\rm ab})] \right\}
\nonumber \\
\gamma_{\pm}^{\rm aa}(\omega_n,\upsilon) & = & 1/\left[\Gamma\: \mp \: i\,\omega_n \cdot (1\:-\:\frac{\upsilon}
{c}) \right] \,
\simeq\, 1/\left[\Gamma\: \mp \: i\,\omega_n\right] \\
\gamma_{\pm}^{\rm bb}(\omega_n,\upsilon) & = & 1/\left[ \Gamma\: \mp \: i\,\omega_n \cdot (1\:-\:\frac{\upsilon}
{c}) \right]\,
\simeq\, 1/\left[\Gamma\: \mp \: i\,\omega_n\right] \nonumber
\end{eqnarray}
We note that in arriving at the above equations we have used the similar approximation as in Section II of Menegozzi \& Lamb (1978)
and explained in more detail in Dinh-V-Trung (2009).

In the frequency domain $\omega_n$, the interaction term $\mathsf{V}\cdot\rho$ can be written as the convolution:
\begin{equation}
\mathsf{V}\cdot\rho\,(\omega_n) = \sum_{q=-\infty}^{q=+\infty}\mathsf{V}(\omega_{\rm n-q})\cdot\rho(\omega_{\rm q})
\end{equation} 
The interaction matrix $\mathsf{V}$ between the electric field $\mathbf{E}$ and the masing molecules having a dipole moment
$\mathbf{d}$ can be calculated following 
Deguchi \& Watson (1990):
\begin{equation}
\mathsf{V} = -\left(\mathbf{E}\,\cdot\,\mathbf{d}\right) = 
{\rm Re}\left\{\sum\limits_{\rm M}^{}{\rm E}_{\rm M}\:
{\rm d}_{\rm -M}\right\} 
= {\rm Re}\left\{\sum\limits_{\rm p}^{}{\rm E}_{\rm p}\:{\rm d}_{\rm -p}\right\}
\simeq \frac{1}{2}\sum\limits_{\rm p}^{}{\rm E}_{\rm p}\:{\rm d}_{\rm -p}
\end{equation}
In the last step we have used the rotating wave approximation, retaining only the positive frequency part of
the electric field $\mathbf{E}$. The values of the spherical components ${\rm d}^{\rm ab}_{\rm p}$ of the dipole moment
matrix between the upper magnetic sub-level ${\rm a}$ and lower level ${\rm b}$ of the 
transition J=1--0 can be easily evaluated as follows (Deguchi \& Watson 1990): 
\begin{eqnarray}
{\rm d}_{p=+1}^{\rm ab} & = & {\rm d}_{M=+1}^{\rm ab}(1\,+\,cos\,\theta)/2 \,+\, i\,{\rm d}_{M=0}^{\rm ab}\,sin\,\theta/\sqrt{2} 
\,-\,{\rm d}_{M=-1}^{\rm ab}(1\,-\,cos\,\theta)/2 \nonumber \\
{\rm d}_{p=-1}^{\rm ab} & = & -{\rm d}_{M=+1}^{\rm ab}(1\,-\,cos\,\theta)/2 \,+\, i\,{\rm d}_{M=0}^{\rm ab}\,sin\,\theta/\sqrt{2}
\,+\,{\rm d}_{M=-1}^{\rm ab}(1\,+\,cos\,\theta)/2
\end{eqnarray}
We also define here the complex conjugate components of the dipole moment:
\begin{equation}
\left({\rm d}_{\rm M}^{\rm ba}\right)^{*} = \left(-1\right)^{\rm M}{\rm d}_{\rm -M}^{\rm ab}
\end{equation}
The interaction matrix written in the frequency domain has the following form:
\begin{equation}
\mathsf{V}_{\rm ab}(\omega_n) = \frac{1}{2}\sum_{\rm p=\pm}\, {\rm d}_{\rm -p}^{\rm ab}\, E_{\rm p}(\omega_n) 
\end{equation}
Therefore from {\bf Eq.}~15 we obtain:
\begin{equation}
\rho_{\rm ab}(\omega_n,\upsilon) = -\frac{i}{2\hbar}\sum_{\rm p=\pm,\,q}\,{\rm d}_{\rm -p}^{\rm ab}\,
E_{\rm p}(\omega_{\rm n-q})\cdot[\rho_{\rm bb}(\omega_{\rm q},\upsilon)\:-
\:\rho_{\rm aa}(\omega_{\rm q},\upsilon)]\cdot\gamma_{+}^{\rm ab}(\omega_n,\upsilon) 
\end{equation}
Consequently, {\bf Eq.}~17 becomes:
\begin{equation}
\mathsf{V}^{\rm ab}\cdot\rho_{\rm ab}^*(\omega_n) = \frac{i}{4\hbar}\sum_{\rm p,p', m, q}\,{\rm d}_{-p}^{\rm ab}
({\rm d}_{-p'}^{\rm ab})^* E_{\rm p}(\omega_{\rm m+n})\cdot E_{\rm p'}^*(\omega_{\rm m-q})\cdot
[{\rho_{\rm bb}}^*(\omega_{\rm q})\: - \:{\rho_{\rm aa}}^*(\omega_{\rm q})]\cdot\gamma_{-}^{\rm ab}(\omega_{\rm m},\upsilon)
\end{equation}
Similar expression for the complex conjugate term:
\begin{equation}
{\mathsf{V}^{\rm ab}}^*\cdot\rho_{\rm ab}(\omega_n) = \frac{i}{4\hbar}\sum_{\rm p,p', m, q}\,({\rm d}_{-p}^{\rm ab}
)^*{\rm d}_{-p'}^{\rm ab}
E_{\rm p}^*(\omega_{\rm m-n})\cdot E_{\rm p'}(\omega_{\rm m-q})\cdot
[\rho_{\rm bb}(\omega_{\rm q})\: - \:\rho_{\rm aa}(\omega_{\rm q})]\cdot\gamma_{+}^{\rm ab}(\omega_{\rm m},\upsilon)
\end{equation}
The elements of density matrix can then be written as follows:
\begin{eqnarray}
\rho_{\rm aa}(\omega_{\rm n},\upsilon)& = & \frac{1}{4\hbar^2} 
\left\{ \sum_{\rm mq\, pp'} {\rm d}_{\rm -p}^{\rm ab} ({\rm d}_{\rm -p'}^{\rm ab})^*
E_{\rm p}(\omega_{\rm m+n})E_{\rm p'}^*(\omega_{\rm m-q})\left[\rho_{\rm bb}^{*}(\omega_{\rm q},
\upsilon)\:-\: \rho_{\rm aa}^{*}(\omega_{\rm q},\upsilon)\right]\gamma_{-}^{\rm ab}(\omega_{\rm m},\upsilon) \right. 
\nonumber \\
& & \left.+\: \sum_{\rm mq\, pp'} ({\rm d}_{\rm -p}^{\rm ab})^* {\rm d}_{\rm -p'}^{\rm ab}
E_{\rm p}^*(\omega_{\rm m-n})E_{\rm p'}(\omega_{\rm m-q})\left[\rho_{\rm bb}(\omega_{\rm q},
\upsilon)\:-\: \rho_{\rm aa}(\omega_{\rm q},\upsilon)\right]\gamma_{+}^{\rm ab}(\omega_{\rm m},\upsilon) 
\: + \right.\nonumber \\ 
 & & \left. \lambda_{\rm a}(\upsilon)\delta_{\rm n,0} 
 \vphantom{\sum_{\rm m}}\right\}\gamma_{+}^{\rm aa}(\omega_{\rm n}) \nonumber \\
\rho_{\rm bb}(\omega_{\rm n},\upsilon)& = & -\frac{1}{4\hbar^2} 
\sum_{\rm a}\left\{ \sum_{\rm mq\, pp'} {\rm d}_{\rm -p}^{\rm ab} ({\rm d}_{\rm -p'}^{\rm ab})^*
E_{\rm p}(\omega_{\rm m+n})E_{\rm p'}^*(\omega_{\rm m-q})\left[\rho_{\rm bb}^{*}(\omega_{\rm q},
\upsilon)\:-\: \rho_{\rm aa}^{*}(\omega_{\rm q},\upsilon)\right]\gamma_{-}^{\rm ab}(\omega_{\rm m},\upsilon) \right. 
\nonumber \\
& & \left.+\: \sum_{\rm mq\, pp'} ({\rm d}_{\rm -p}^{\rm ab})^* {\rm d}_{\rm -p'}^{\rm ab}
E_{\rm p}^*(\omega_{\rm m-n})E_{\rm p'}(\omega_{\rm m-q})\left[\rho_{\rm bb}(\omega_{\rm q},
\upsilon)\:-\: \rho_{\rm aa}(\omega_{\rm q},\upsilon)\right]\gamma_{+}^{\rm ab}(\omega_{\rm m},\upsilon) \right. 
\nonumber \\
 & & \left. + \: \lambda_{\rm b}(\upsilon)\delta_{\rm n,0} 
 \vphantom{\sum_{\rm m}}\right\}\gamma_{+}^{\rm mm}(\omega_{\rm n})
\end{eqnarray}
The polarization vector $\mathbf{P}$ of the masing medium induced by the radiation field is calculated as:
\begin{equation} 
\mathbf{P}(z,t) = \int\limits_{-\infty}^{\infty}d\upsilon\, \mathbf{tr}
[\rho(z,\upsilon,t)\, \mathbf{d}]
\end{equation} 
or written explicitly using the rotating wave approximation:
\begin{equation}
\frac{1}{2}P_{p}(z,\omega_n) \simeq \int_{-\infty}^{\infty} \, d\upsilon \, 
\sum\limits_{\rm a}^{}\rho_{\rm ab}(z,\upsilon,\omega_n)\,{\rm d}^{\rm ba}_{\rm p}
\end{equation}
Once the equations of the density matrix components are solved in the frequency domain, we can
calculate the polarization vector of the maser medium using the above expression.\\
Using the normalised homogeneous line profile $\phi^{\rm ab}_\pm$ defined as follows:
\begin{equation}
\phi^{\rm ab}_\pm(\omega_n,\upsilon) = \frac{\omega_0}{\pi {\rm c}}\cdot\gamma_{\pm}^{\rm ab}(\omega_n,\upsilon)
\end{equation}
the transfer equations {\bf Eqs.}~11 for maser radiation become:
\begin{equation}
\frac{dE_{\rm p}(z,\omega_n)}{dz} = \frac{2\pi^2}{\hbar}\int d\upsilon \sum_{\rm a p' q}
({\rm d}_{\rm -p}^{\rm ab})^* {\rm d}_{\rm -p'}^{\rm ab}E_{\rm p'}(z,\omega_{n-q}) \left[\rho_{\rm aa}(z,\upsilon,\omega_q)\: -
\: \rho_{\rm bb}(z,\upsilon,\omega_q)\right]\phi_{+}^{\rm ab}(\omega_{\rm n},\upsilon) 
\end{equation}
These equations are of great importance in our work because they govern the change of the amplitudes and phases of 
the background radiation through the maser medium once the density matrix components are known. In the next section
we will use these equations to follow the evolution of the amplitudes and phases of different Fourier components of the
radiation field through the maser medium
\subsection{First order approximation: radiation transfer equations for Stokes parameters}
In order to derive the standard transfer equations for maser radiation, we need to make two assumptions: the
population inversion is constant in time ($\rho_{\rm aa}(\omega_{\rm q},\upsilon)$=0 for $\omega_{\rm q} \neq 0$) 
and the Fourier components of the radiation field at different frequencies are not correlated. In addition, 
astronomical masers are known to be broadband, i.e the line width due to Doppler broadening is 
much larger than the homogeneous linewidth $\Gamma$ and the transition rate due to stimulated emission. 
Thus we can safely ignore the imaginary part of the homogeneous line profile $\gamma^{\rm ab}_\pm$.   
After straightforward manipulations of equations {\bf Eqs.}~25 and {\bf Eqs.}~29, and taking the ensemble average,
we obtain the following form of the
transfer equations for the Stokes parameters of the radiation field:
\begin{equation}
\frac{d}{dz}\left( \begin{array}{l} I(\omega) \\ Q(\omega) \\ U(\omega) \\ V(\omega) \end{array}\right) =
\left( \begin{array}{cccc}
A(\omega) & B(\omega) & 0 & C(\omega) \\
B(\omega) & A(\omega) & 0 & 0 \\
0 & 0 & A(\omega) & 0 \\
C(\omega) & 0 & 0 & A(\omega) \end{array}\right)\: \left( \begin{array}{l} I(\omega) \\ Q(\omega) \\ 
U(\omega) \\ V(\omega) \end{array}\right)
\end{equation}
The coefficients A($\omega$), B($\omega$) and C($\omega$) are defined as follows:
\begin{eqnarray}
A(\omega) & = &\frac{h\nu}{4\pi}\cdot 3\,B\cdot\frac{c}{\nu}
\:\cdot\frac{1}{4}\left[\left(\Delta\rho_{\rm ++}\, +\, 
\Delta\rho_{\rm --}\right)\left(1\;+\;{\rm cos^2}\,\theta\right)\;+\;
2\,\Delta\rho_{\rm 00}\,{\rm sin}^{\rm 2}\,\theta\right] \nonumber \\
B(\omega) & = &\frac{h\nu}{4\pi}\cdot 3\,B\cdot\frac{c}{\nu} 
\:\cdot\frac{1}{4}\left[\left(\Delta\rho_{++}\;+\;\Delta\rho_{--}\right)
\;-\;2\,\Delta\rho_{\rm 00}\right]{\rm sin^2}\,\theta \\
C(\omega) & = &\frac{h\nu}{4\pi}\cdot 3\,B\cdot\frac{c}{\nu}
\:\cdot\frac{1}{2}\left[\Delta\rho_{++}\;-\;\Delta\rho_{--}
\right]\,{\rm cos}\,\theta \nonumber
\end{eqnarray}
In the above equations, $B$ is the Einstein coefficient and related to the spontaneous emission probability $A$ by the
well-known relation $B\:=\:A\cdot c^2/2h\nu^3$ and 
\begin{eqnarray}
\Delta\rho_{\rm ++} & = & \int_{-\infty}^{+\infty}\:d\upsilon\:\phi^{\rm +b}_{\rm r}(\omega,\upsilon)
[\rho_{\rm ++}(\upsilon)\,-\,\rho_{\rm bb}(\upsilon)] \nonumber \\
\Delta\rho_{\rm 00} & = & \int_{-\infty}^{+\infty}\:d\upsilon\:\phi^{\rm 0b}_{\rm r}(\omega,\upsilon)
[\rho_{\rm 00}(\upsilon)\,-\,\rho_{\rm bb}(\upsilon)] \\
\Delta\rho_{\rm --} & = & \int_{-\infty}^{+\infty}\:d\upsilon\:\phi^{\rm -b}_{\rm r}(\omega,\upsilon)
[\rho_{\rm --}(\upsilon)\,-\,\rho_{\rm bb}(\upsilon)] \nonumber
\end{eqnarray}
where $\phi^{\rm ab}_{\rm r}(\omega,\upsilon)$ is the real part of the normalised homogeneous line profile
$\phi^{\rm ab}_\pm(\omega,\upsilon)$.
The density matrix representing population in each magnetic sub-level of the masing molecules is
determined through the familiar statistical equilibrium equations:
\begin{eqnarray}
\Gamma\rho_{\rm ++}(\upsilon) &=& -R_{\rm ++}[\rho_{\rm ++}(\upsilon)\:-\:\rho_{\rm bb}(\upsilon)] 
\,+\, \lambda_{+}(\upsilon)\nonumber \\ 
\Gamma\rho_{\rm 00}(\upsilon) &=& -R_{\rm 00}[\rho_{\rm 00}(\upsilon)\:-\:\rho_{\rm bb}(\upsilon)] 
\,+\, \lambda_{0}(\upsilon)\\
\Gamma\rho_{\rm --}(\upsilon) &=& -R_{\rm --}[\rho_{\rm --}(\upsilon)\:-\:\rho_{\rm bb}(\upsilon)] 
\,+\, \lambda_{-}(\upsilon)\nonumber  \\
\Gamma\rho_{\rm bb}(\upsilon) &=& R_{\rm ++}[\rho_{\rm ++}(\upsilon)\:-\:\rho_{\rm bb}(\upsilon)] + 
R_{\rm 00}[\rho_{\rm 00}(\upsilon)\:-\:\rho_{\rm bb}(\upsilon)] + \nonumber \\  
 & & R_{\rm --}[\rho_{\rm --}(\upsilon)\:-\:\rho_{\rm bb}(\upsilon)] 
\,+\, \lambda_{\rm b}(\upsilon)\nonumber  
\end{eqnarray} 
where the stimulated emission rate $R_{\rm ++}$, $R_{\rm 00}$ and $R_{\rm --}$ are defined as:
\begin{eqnarray}
R_{\rm ++} &=& -\frac{3B}{8\pi^2}\sum_{\rm m}[\frac{1\:+\:cos^2\theta}{2}I(\omega_{\rm m})\: + \: 
cos\theta\,V(\omega_{\rm m}) \:+\:\frac{sin^2\theta}{2}Q(\omega_{\rm m})]
\gamma^{\rm +b}_{\rm r}(\omega_{\rm m},\upsilon)\Delta\omega \nonumber \\ 
R_{\rm 00} &=& -\frac{3B}{8\pi^2}\sum_{\rm m}\,sin^2\theta[I(\omega_{\rm m})\:-\:Q(\omega_{\rm m})]
\gamma^{\rm 0b}_{\rm r}(\omega_{\rm m},\upsilon)\Delta\omega \\
R_{\rm --} &=& -\frac{3B}{8\pi^2}\sum_{\rm m}[\frac{1\:+\:cos^2\theta}{2}I(\omega_{\rm m})\: - \: 
cos\theta\,V(\omega_{\rm m}) \:+\:\frac{sin^2\theta}{2}Q(\omega_{\rm m})]
\gamma^{\rm -b}_{\rm r}(\omega_{\rm m},\upsilon)\Delta\omega \nonumber 
\end{eqnarray} 
In the above expressions, the functions $\gamma^{\rm ab}_{\rm r}(\omega,\upsilon)$ are the real part of
the homogeneous line profiles $\gamma^{\rm ab}_\pm(\omega,\upsilon)$.\\
Assuming a small Zeeman splitting and using the fact that the functions 
$\gamma^{\rm ab}_{\rm r}(\omega,\upsilon)$ are sharply peaked in comparison to
the maser linewidth and have the normalisation 
$\sum_{\rm m}\,\gamma^{\rm ab}_{\rm r}(\omega_{\rm m},\upsilon)\Delta\omega\,=\,\pi$, 
we can easily recover the standard equations derived by Goldreich et al. (1973), Deguchi \& Watson (1990).
Therefore, the standard radiative transfer equations for the Stokes parameters follow naturally from
our formulation of the astronomical maser.
\section{Simulations of maser amplification}
In our simulation we will choose the parameters appropriate to an astronomical maser.
For simplicity, we choose the loss rate $\Gamma$ = 1 s$^{-1}$ and the normalized pump rates are assumed to have
a velocity dispersion $\sigma$, with the difference in pump rates $\Delta\lambda(\upsilon)\,=\,
\exp(-\upsilon^2/\sigma^2)$. In our simulation we use 600 modes with a frequency resolution of 
$\Delta\omega=0.75\,\Gamma$ around the maser line covering the range
$-\sigma$ to $+\sigma$ in the velocity domain. 
The actual velocity resolution $\Delta\upsilon$ 
will depend on the frequency of the maser line, as $\Delta\upsilon$ = $[c/\omega_{0}]\Delta\omega$.
For a maser line such as the 1612 MHz OH maser, $\Delta\upsilon$ is approximately 2 cm s$^{-1}$. The
corresponding value for the velocity dispersion is 200 cm s$^{-1}$
Although the velocity dispersion of the maser line in our simulations is much smaller than in real 
astronomical masers, the number of frequency modes and the bandwidth are large enough, i.e. the bandwidth
of 150 s$^{-1}$ is much greater than the loss rate of 1 s$^{-1}$,
to capture the main features of the broadband radiation field produced by the astronomical masers.\\
The Zeeman splitting of the upper energy level $J=1$ is taken into account explicitly in our simulations. We
adopt a value of the splitting $g\omega_B$ = $10\,\Delta\omega$ or 7.5 s$^{-1}$. The adopted splitting is consistent
with our starting assumption of the presence of a strong B-field in the maser because $g\omega_B$ is much larger than
the loss rate $\Gamma$ = 1 s$^{-1}$ from the energy levels involved in the maser transition. \\
Since we deal with only a partially saturated maser, we consider
only 10 harmonic components of the density matrix $\rho_{\rm aa,\; bb}(\omega_{\rm q}, \upsilon)$. As shown later, 
the number of harmonic components
is enough to capture the pulsations of the molecular population inversion.
We generate the background continuum radiation in a simimlar way as Menegozzi \& Lamb (1978) 
using random generator RAN2 from Press et al. (1992). The phase of
electric field components is random and uniformly distributed over the interval 0 to 2$\pi$. 
To reduce the enormous amount of computer time required to compute enough realisations of the radiation
field in order to reach an acceptable level of random noise in the Stokes parameters, 
we choose to use a constant amplitude for all the modes of the background radiation field. Our choice of
course allows us to study only the amplification of the background radiation field with radom phases. However,
the difference in phases between frequency modes of the radiation field is enough to randomize the Stokes
parameters ${\cal Q}$ and ${\cal U}$, which characterize the linear polarization property of the background 
radiation field. Therefore, in our simulations the amplitude of all frequency modes is a constant 
$I_\pm(\omega_{\rm n})\Delta\omega\,=\,$1 on scale of $2h\nu^3/c^2$. 
The intensity of the maser shown in all figures is also of the form $I(\omega_{\rm n})\Delta\omega$, where
$I(\omega_{\rm n})$ is the intensity per unit frequency $\nu$ as defined in the previous section.  
We use a fourth-order Runge-Kutta method with a fixed 
step $h\,=\,$0.02 to integrate the transfer equations {\bf Eqs.}~29.
Each realisation of the background radiation field is evolved through the maser 
by solving the equations {\bf Eqs.}~25 to determine the density matrix, together with 
integrating the equations {\bf Eq.}~29 to calculate the change of the amplitudes and phases of different
spherical components ($E_{-}$ and $E_+$) of the radiation field. 
We then record the emergent radiation for later analysis. For the sake of simplicity, we will present
the results in the form of the parameters $({\cal I,Q,U,V})$, as defined in Sec.2.1 for each realisation of
the background continuum radiation. These
parameters have the advantage of containing the same information on the amplitudes and phases as the 
spherical components $E_{-}$ and $E_+$, and at the same time are directly related to the usual Stokes parameters
through a simple ensemble average.\\
We carry out our simulations in the partially saturated regime, which is likely relevant to most astronomical
masers. The choice is also necessary because we consider only a limited number of harmonic components of the
molecular population inversion. In the saturated regime the fluctuation will be stronger and thus require 
the consideration of a larger number of harmonic components. The amplification of the maser with length $L$ is specified 
by the unsaturated optical depth $\tau(\upsilon\,=\,0)$ at the line center:
\begin{equation}
\tau(\upsilon) = \frac{h\nu}{4\pi}\cdot 3\,B\cdot\frac{c}{\nu}\,\frac{\Delta\lambda(\upsilon)}{\Gamma}\,L
\end{equation}
A value of $\tau(\upsilon\,=\,0)$ = 20 is used throughout in our
simulations. The spontaneous transition rate between the upper and lower maser levels is 
taken to be 10$^{-9}$ s$^{-1}$.
The Einstein coefficient $B$ is related to the spontaneous transition rate by the 
well-known relation $A\,=\,[2h\nu^3/c^2]\,B$.
Our choice of unsaturated optical depth corresponds to a partially saturated maser. As such, the number
of harmonic components of the density matrix used in our simulation is adequate to capture the pulsations
induced by the radiation field. However, for higher optical depth, a larger number of harmonic components will be
necessary.\\
We consider two representative cases: in the first case the magnetic field $\mathbf{B}$ makes an angle $\theta\:=\:30^0$ 
 with respect
to the maser axis and in the second case the magnetic field is perpendicular ($\theta\:=\:90^0$) to the maser axis.
The partially saturated masers in these two cases allow us to assess the validity of the standard formulation and  also
test the predictions of the model presented in Elitzur (1992, 1993, 1996).\\
In {\bf Fig.}~1 we show the parameters ($\mathcal{I}$, 
$\mathcal{Q}$, $\mathcal{U}$, $\mathcal{V}$) of one realisation of the background continuum radiation for the 
case $\theta\:=\:30^0$. Because in our simulations 
the amplitude of both $E_{-}$ and $E_{+}$ of quasi-monochromatic modes are the same, 
the parameter $\mathcal{V}$ is identical to zero. The intensity $\mathcal{I}$ is also 
constant across the bandwidth of the incident radiation.
Only the parameters $\mathcal{Q}$ and $\mathcal{U}$ of the radiation field fluctuate strongly
due to the random phase of the electric field components $E_{+}$ and $E_{-}$.
However, the ensemble average Stokes parameters Q and
U are identically zero, as expected from a non-polarized background continuum radiation.
The amplified radiation at the output of the maser as shown in {\bf Fig.}~2 clearly displays the line
narrowing effect and intensity fluctuations induced by the population pulsations. Because the maser
is only partially saturated, the harmonic components $\rho(z,\omega_q,\upsilon)$ of the magnetic sub-level populations are 
small for $\omega_q\:\neq\:0$ as seen in {\bf Fig.}~3.
In {\bf Fig.}~4 we show the ensemble averaged Stokes parameters of the maser radiation calculated using 2400 
realisations of the background continuum radiation field. These Stokes parameters are consistent with the 
results of standard formulation for the same case as shown in {\bf Fig.}~5. 
Similar results for the case $\theta\:=\:90^0$ are shown in {\bf Fig.}~6 and 7.  
\section{Discussion}
Our simulation results show that the amplified radiation propagating close to the direction of the magnetic
field ($\theta\:=\:30^0$) possesses both linear and circular polarization. 
The appearance of circular polarization is a natural consequence of taking into account the Zeeman splitting, which in our case 
is not negligible in comparison to the maser linewidth. Since the stimulated emission rate is quite small
even at the line centre ($R_{\pm\pm}/\Gamma\:\sim\:$0.05, $R_{00}/\Gamma\:\sim\:$0.1), 
the fractional linear polarization Q/I is small $\sim\:-$1\% at the line center, well below the limit of 100\% predicted
by Goldreich et al. (1973) for fully saturated masers. Within the partial saturation regime, which is probably 
most relevant to astronomical masers, our results are consistent with the results obtained from standard formulation 
({\bf Eqs.}~30 and {\bf Eqs.}~33) and with the calculations of Western \& Watson (1984), which indicate that 
linear polarization approaches the 100\% limit very slowly for $\sin^2\theta\:\le\:1/3$. 
The fractional linear polarization of the maser polarization grows faster in the case $\theta\:=\:90^0$ and 
reaches Q/I $\sim$1.5\% at the line center ({\bf Figs.}~6 \& 7). 
This result is also consistent with that predicted by Western \& Watson (1984). We note that by reducing the multi-level model
to the idealized two-level case, Gray (2003) also reached similar conclusions.
Thus our results agree with previous analytical and numerical simulation works done by Gray (2003) and by Western \& Watson (1984).\\
The creation of linear polarization when the magnetic field is close to the maser axis 
($\theta\,=\,30^0$ or $\sin^2\theta\,=\,1/4$)
contradicts the model of maser polarization described in Elitzur (1993, 1996), which predicts that
propagation of polarized radiation is inhibited in the directions close to the magnetic field {\bf B}. 
No instability associated with the evolution of fully polarized modes of maser radiation is 
seen during the integration of the radiation transfer equations {\bf Eqs.}~25
and {\bf Eqs.}~29. Such instability is identified in Elitzur (1993) as the main reason for maser radiation to 
be linearly polarized at the polarization limit Q/I = (3$sin^2\,\theta - 2$)/3$sin^2\,\theta$ for 
$sin^2\,\theta\,\ge \, \frac{1}{3}$ even when the maser is unsaturated. Contrary to this assertion, our simulation 
for the case of a partially saturated maser with $\theta\:=\:90^0$ shows that the fractional linear polarization of
the maser radiation is actually small, $\sim$1.5\%, well below the limit of 33\% given by the above expression.\\ 
We note that in Elitzur (1993, 1996) the equations normally reserved
for Stokes parameters and steady-state populations, quantities obtained by ensemble averaging over large number of realisations of
radiation fields, are used to describe the evolution of quasi-monochromatic and fully polarized modes of the radiation field. 
Consequently, it leads to some difficulty in understanding the requirement given in Elitzur (1993, 1996)
to perform another averaging step to justify the mathematical solutions of 
the eigenvalue problem for the standard radiation transfer equations. The confusing description of the way to handle the random 
radiation field (Elitzur 1991, 1993) seems to indicate that a self-consistent treatment of the interaction between radiation and 
masing medium has not been achieved.    
\section{Conclusion}
The explicit incorporation of broadband random radiation field into our treatment of the astronomical maser has proved very 
important to investigate the properties of maser emission. We are able to directly simulate the amplification of the
background continuum radiation by the maser and study in detail the appearance of linear polarization. Our simulation 
results show that there is no problem with previous numerical studies of the maser polarization in the 
unsaturated and partially saturated regime. Hopefully, the current formulation can be extended and applied to study maser emission 
under conditions other than that considered here.  

\section*{Acknowledgments}

We would like to thank the referee, Dr. M.D. Gray, for insightful and constructive comments that help to improve
greatly the presentation of our paper. This research has made use of 
NASA's Astrophysics Data System Bibliographic Services
and the SIMBAD database, operated at CDS, Strasbourg, France.

\newpage
\begin{figure}
\includegraphics[width=8cm]{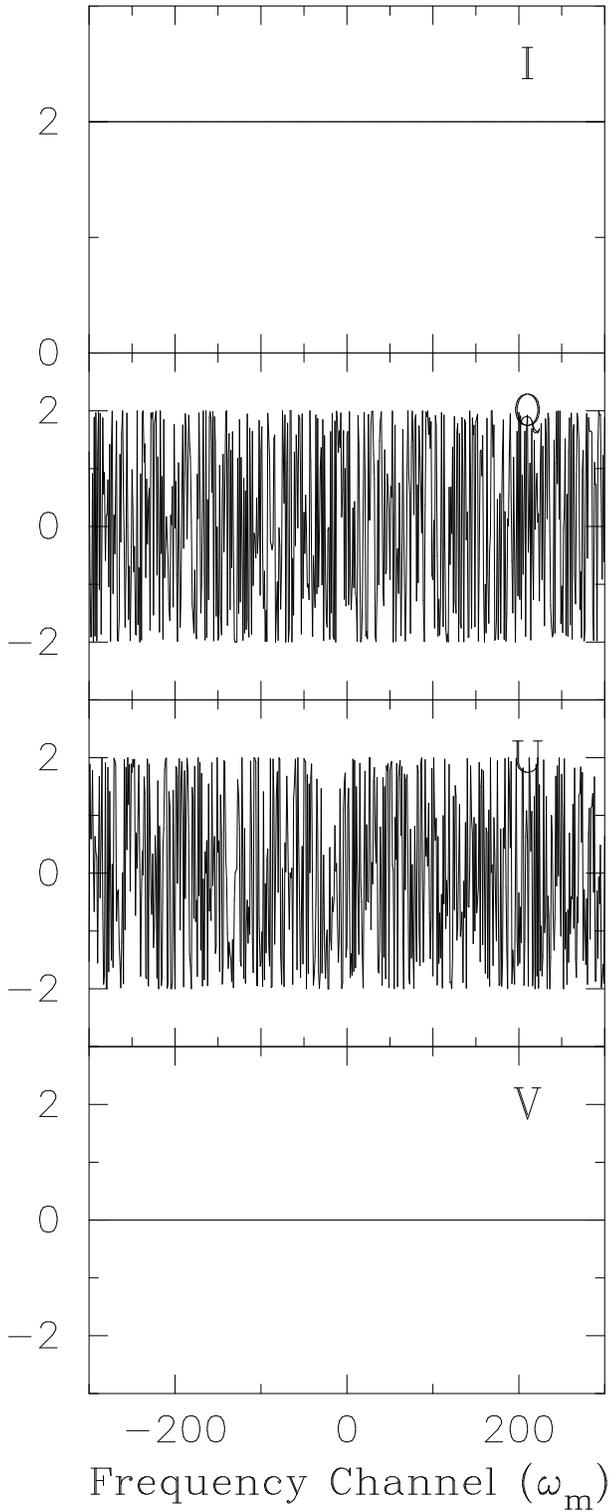}
\caption{One realisation of the background radiation field. The Stokes parameters ($\mathcal{I}$, 
$\mathcal{Q}$, $\mathcal{U}$, $\mathcal{V}$) of the quasi-monochromatic modes across the bandwith of 
the radiation field are shown in each frame. All the parameters $({\cal I,Q,U,V})$ are expressed on 
a scale of 2$h\nu^3/c^2$. The Stokes parameters I and V have constant values because of the assumption
on the constant amplitude of the frequency modes of the background continuum radiation.}
\label{fig1}
\end{figure}

\begin{figure}
\includegraphics[width=8cm]{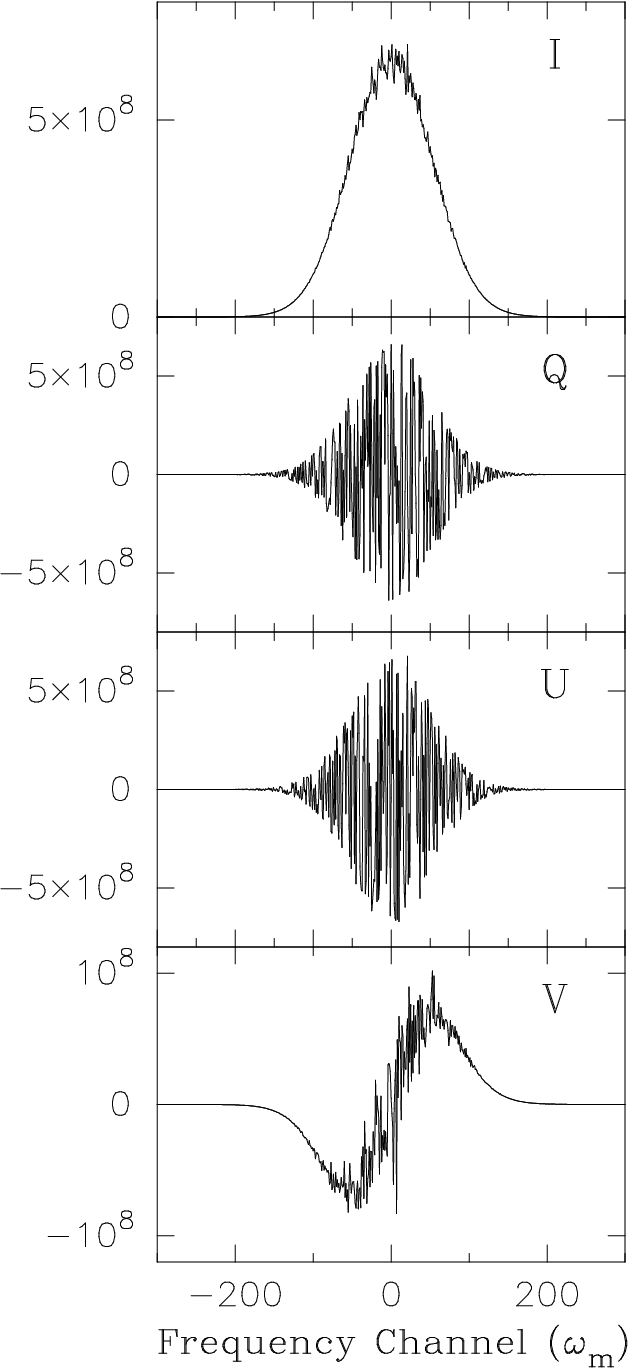}
\caption{The emergent radiation field amplified by the maser, corresponding to the incident radiation field shown in
Fig.1. The magnetic field makes an angle $\theta$=30$^0$ with respect to the maser axis. 
The Stokes parameters ($\mathcal{I}$, 
$\mathcal{Q}$, $\mathcal{U}$, $\mathcal{V}$) of the quasi-monochromatic modes across the maser line profile are shown
in each frame.
All the parameters $({\cal I,Q,U,V})$ are expressed on a scale of 2$h\nu^3/c^2$.}
\label{fig2}
\end{figure}

\begin{figure}
\includegraphics[width=8cm]{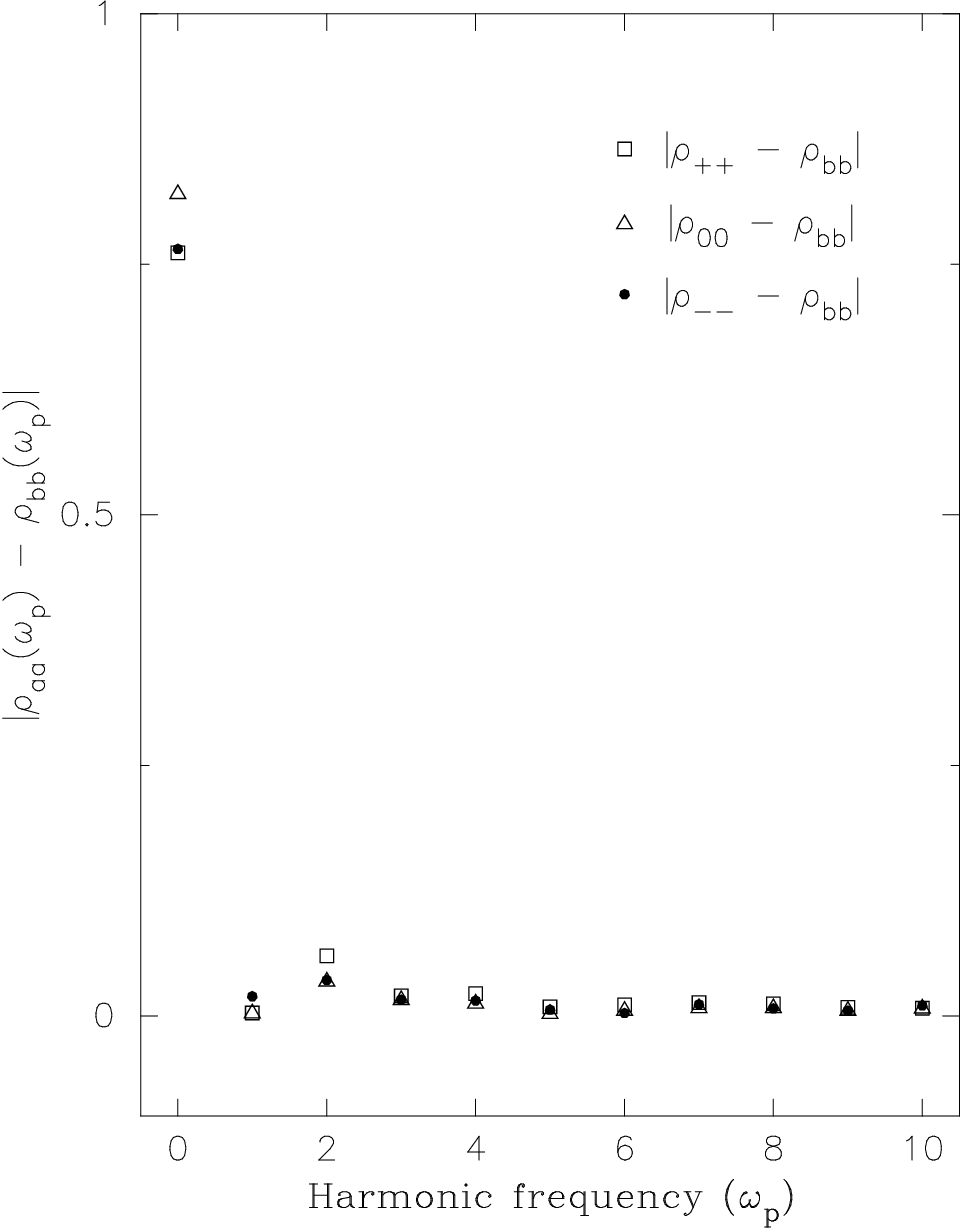}
\caption{Harmonic components of the magnetic sub-level populations of molecules with zero velocity ($\upsilon$=0) at
the end point of the maser ($z\:=\:L$). The corresponding emergent radiation field is shown in Figure. 2}
\label{fig3}
\end{figure}

\begin{figure}
\includegraphics[width=8cm]{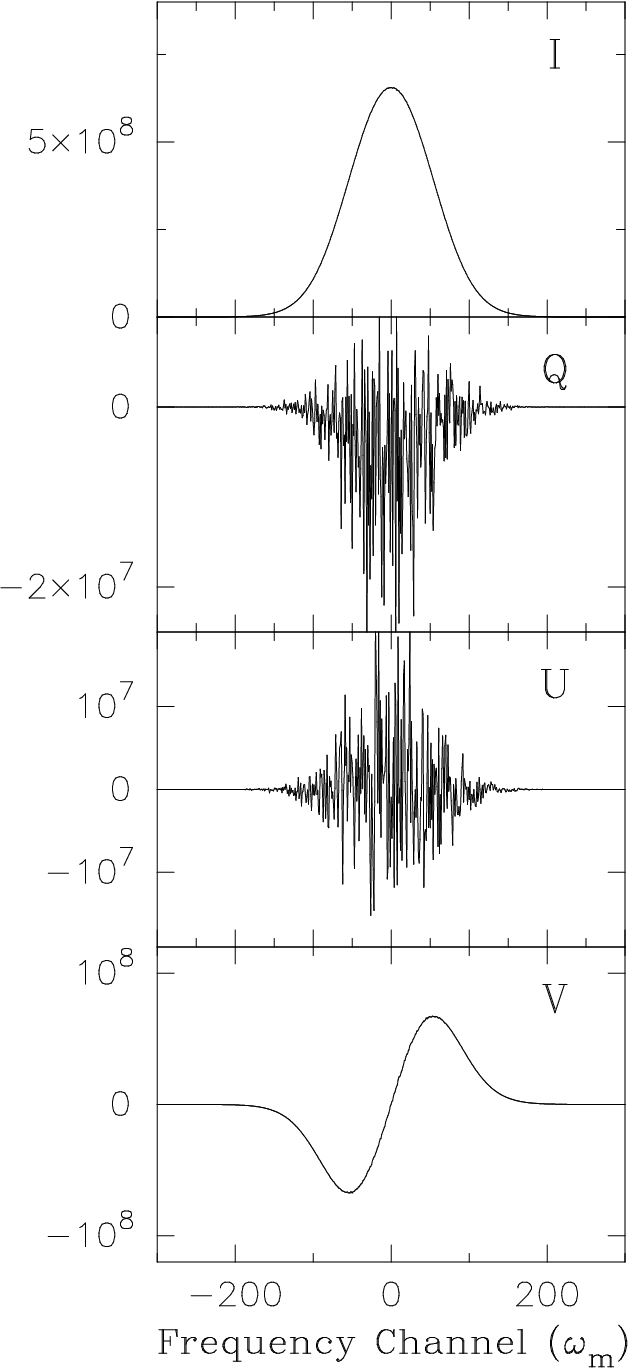}
\caption{Stokes parameters of the maser radiation field determined from the ensemble average of the
parameters $({\cal I,Q,U,V})$ over 2400 realisations
of the radiation fields. The magnetic field is inclined at an angle $\theta\:=\:30^0$ 
with respect to the maser axis. The Stokes parameters are expressed on a scale of 2$h\nu^3/c^2$.}
\label{fig4}
\end{figure}

\begin{figure}
\includegraphics[width=8cm]{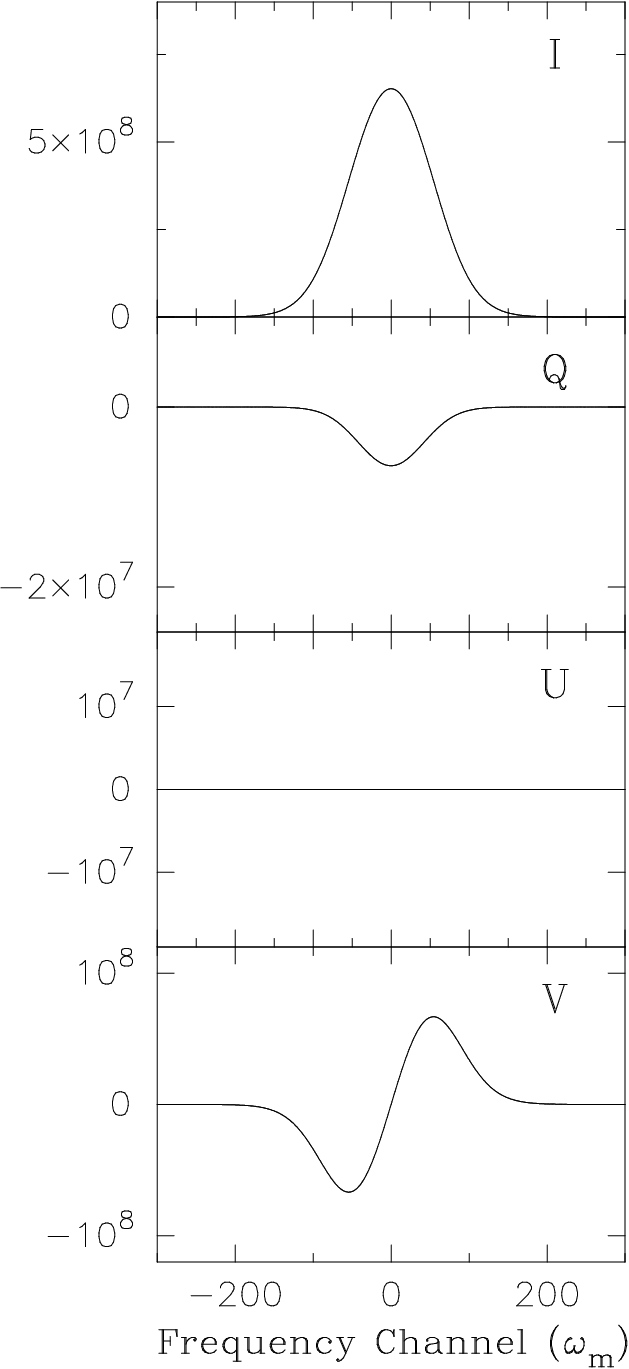}
\caption{Stokes parameters of the maser radiation field predicted using the standard formulation in the
case $\theta\:=\:30^0$. The Stokes parameters are expressed on a scale of 2$h\nu^3/c^2$.}
\label{fig5}
\end{figure}

\begin{figure}
\includegraphics[width=8cm]{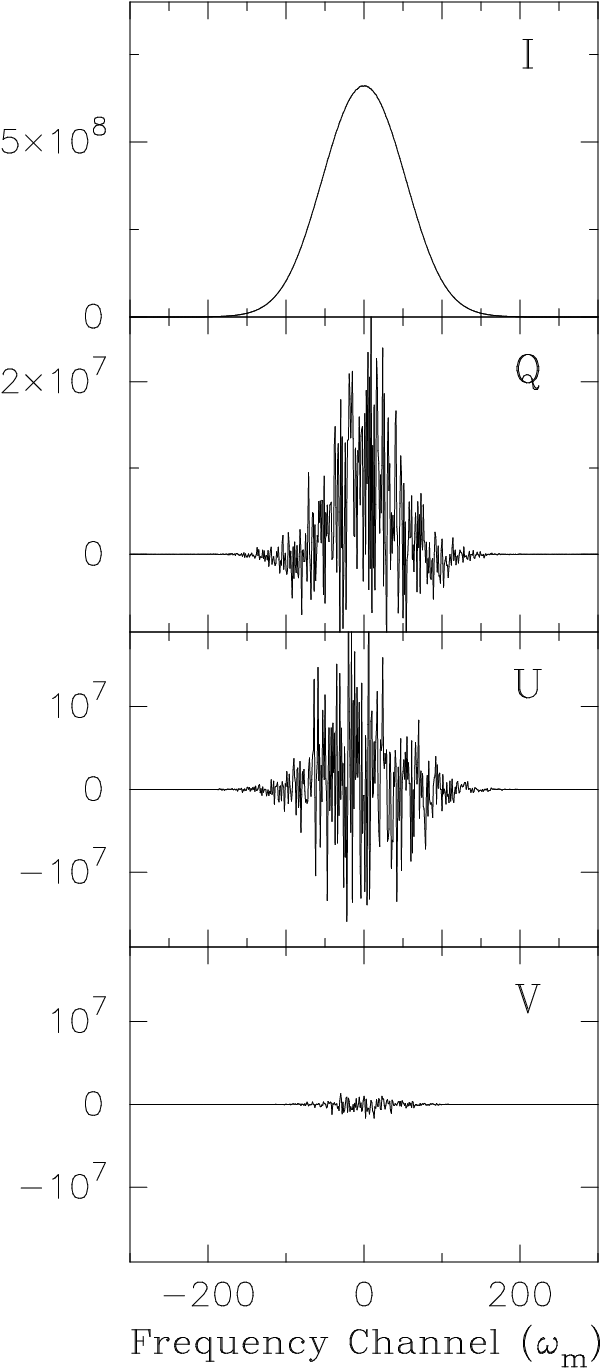}
\caption{Stokes parameters of the maser radiation field determined from the ensemble average of the
parameters $({\cal I,Q,U,V})$ over 2400 realisations
of the radiation field. The magnetic field is inclined at an angle $\theta\:=\:90^0$ 
with respect to the maser axis. The Stokes parameters are expressed on a scale of 2$h\nu^3/c^2$.}
\label{fig6}
\end{figure}

\begin{figure}
\includegraphics[width=8cm]{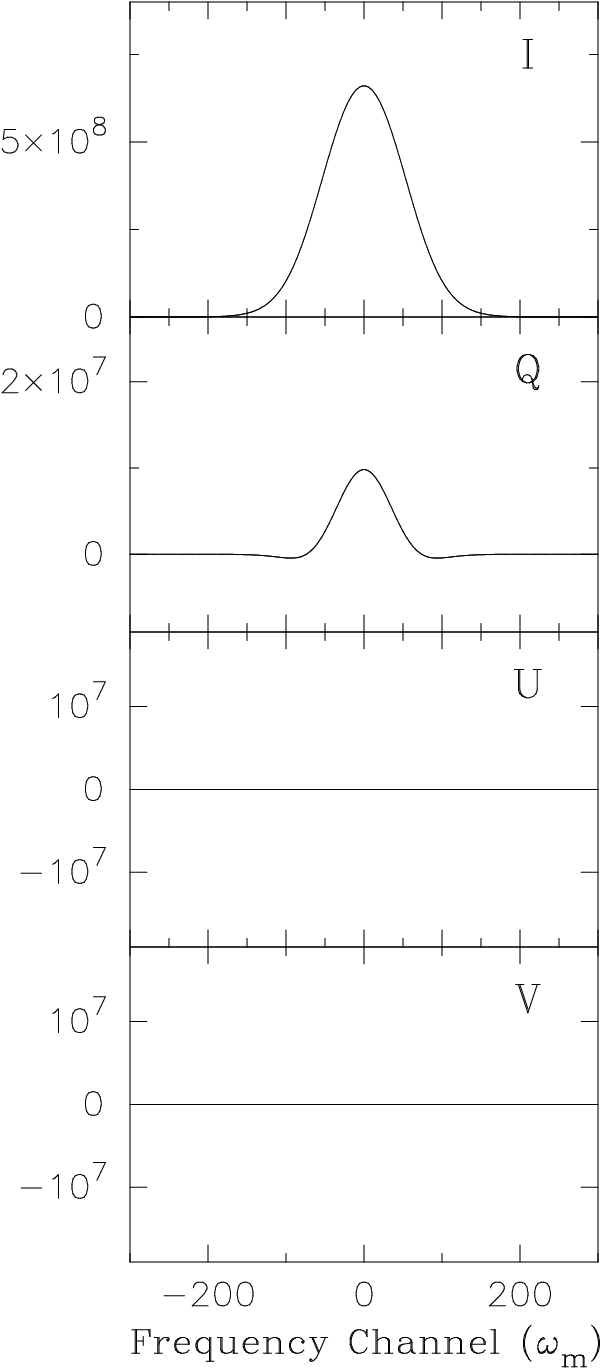}
\caption{Stokes parameters of the maser radiation field predicted using the standard formulation in
the case $\theta\:=\:90^0$. The Stokes parameters are expressed on a scale of 2$h\nu^3/c^2$.}
\label{fig7}
\end{figure}
\end{document}